\documentclass[twocolumn]{jpsj2} 
\title{Kondo problems in Tomonaga-Luttinger liquids}

\author{Akira \textsc{Furusaki}}

\inst{Condensed Matter Theory Laboratory, RIKEN, Wako, Saitama 351-0198}

\abst{
Quantum impurity problems in Tomonaga-Luttinger liquids (TLLs) are
reviewed with emphasis on their analogy to the Kondo problem
in Fermi liquids.
First, the problem of a static impurity in a spinless TLL
is considered, which is related to the model studied in
the context of the macroscopic quantum coherence.
In the low-energy limit the TLL is essentially
cut into two pieces when interaction is repulsive.
The orthogonality catastrophe in a TLL is then discussed.
Finally, the Kondo effect of a spin-1/2 impurity in a one-dimensional
repulsively interacting electron liquids (a spinful TLL) is reviewed.
Regardless of the sign of the exchange coupling, the impury spin is
completely screened in the ground state.
The leading low-temperature contributions to thermodynamic quantities
come from boundary contributions of a bulk leading
irrelevant operator.
}

\kword{Kondo effect, Tomonaga-Luttinger liquid,
 bosonization, orthogonality catastrophe}

\begin{document}
\maketitle

\section{Introduction}

One-dimensional strongly interacting electrons have been a very
actively studied subject for decades.
In one spatial dimension mutual interactions among electrons have
dramatic impact on low-energy properties of electron liquids
and change them into Tomonaga-Luttinger liquids (TLLs)
in which low-energy excitations are not
fermionic single-particle excitations but
bosonic gapless collective excitations of density fluctuations.
The defining feature of a TLL is continuously varying
exponents of correlation functions in its ground state,
which experimentally manifests itself as power-law
temperature dependence of response functions.
For example, conductance of a junction formed in a TLL exhibits
a power-law temperature dependence reflecting the energy
dependence of tunnel density of states with an exponent
depending on interaction strength.
This characteristic feature is observed successfully in experiments of
quantum wires,\cite{tarucha} edge states in fractional
quantum Hall liquids,\cite{chang} and carbon nanotubes.\cite{bockrath}

Along with these experimental developments, quantum impurity problems
in TLLs have been intensively studied theoretically.
A minimal model among those studied is a spinless TLL with a
static impurity potential.\cite{kane92prl}
This model is closely related to a problem studied
intensively in the context of macroscopic quantum coherence:
quantum mechanics of a particle moving in a periodic potential
with Ohmic dissipation.
When the periodic potential is replaced by a double-well potential,
the model is known to be equivalent to the Kondo
problem.\cite{leggett}
Moreover, at a particular value of a TLL parameter
$g=\frac{1}{2}$, it is equivalent to the two-channel
Kondo problem.\cite{matveev95}
An interesting generalization of these models is the one of
a spinful TLL with a spin-$\frac{1}{2}$ magnetic impurity.
A purpose of this paper is to review this Kondo problem
in a TLL.\cite{dhlee,furusaki94,frojdh,durganandini}
In view of the above-mentioned similarity, we first review the
static impurity problem (which will then serve as a basis for
understanding the Kondo problem),
the Anderson's orthogonality catastrophe in a TLL,
and finally the Kondo problem.
Since this paper is a short article, we cannot
cover many important developments such as exact results obtained
through the Bethe ansatz.
Instead, we will concentrate on low-energy effective field
theory approach using the Abelian bosonization method.

\section{A static impurity}

\subsection{spinless case}

The starting point of our discussion is a simple continuum model
of interacting spinless fermions with linear dispersion.
The model can be applied to quantum wires in strong magnetic fields
and to edge states in principal fractional quantum Hall liquids.
Its Hamiltonian is written as
\begin{align}
\mathcal{H}_0 & =
\int^\infty_{-\infty}\!dx \left\{
iv_F\left[\psi^\dagger_L(x)\partial_x\psi^{}_L(x)
-\psi^\dagger_R(x)\partial_x\psi^{}_R(x)\right]
\right.
\nonumber\\
& \quad
\left.{}
+U\left[:\psi^\dagger_L(x)\psi^{}_L(x):
       +:\psi^\dagger_R(x)\psi^{}_R(x):\right]^2
\right\},
\end{align}
where $v_F$ is Fermi velocity, $\psi_L(x)$ and $\psi_R(x)$ are
annihilation operators of left- and right-going fermions.
We have discarded umklapp scattering, assuming that
particle density is away from half filling.
To simplify the Hamiltonian, we bosonize the fermions,
\begin{equation}
\psi^{}_R(x)=\frac{e^{i\varphi_R(x)}}{\sqrt{2\pi\alpha}},
\quad
\psi^{}_L(x)=\frac{e^{-i\varphi_L(x)}}{\sqrt{2\pi\alpha}},
\end{equation}
where the bosonic fields obey the commutation relations
\begin{subequations}
\begin{align}
&
[\varphi_R(x),\varphi_R(y)]=-[\varphi_L(x),\varphi_L(y)]
=i\pi\mathrm{sgn}(x-y),\\
&
[\varphi_R(x),\varphi_L(y)]=i\pi,
\end{align}
\end{subequations}
and $\alpha$ is a short-distance cutoff.
The particle density is written as
\begin{subequations}
\begin{align}
&
:\!\psi^\dagger_R(x)\psi^{}_R(x)\!:\,=\frac{1}{2\pi}\frac{d\varphi_R}{dx},
\\
&
:\!\psi^\dagger_L(x)\psi^{}_L(x)\!:\,=\frac{1}{2\pi}\frac{d\varphi_L}{dx},
\end{align}
\end{subequations}
where the fermion operators are normal ordered with respect to the
Dirac sea.
The Hamiltonian is then bosonized as
\begin{equation}
\mathcal{H}_0=\frac{v}{8\pi}\int^\infty_{-\infty}\! dx
\left[\frac{1}{g}\left(\frac{d\phi}{dx}\right)^2
+g\left(\frac{d\theta}{dx}\right)^2\right],
\label{H_0}
\end{equation}
where $g=[1+(2U/\pi v_F)]^{-1/2}$, $v=v_F/g$,
$\phi=\varphi_L+\varphi_R$, and $\theta=\varphi_L-\varphi_R$.
The field $\phi$ represents particle density while
the field $\theta$ corresponds to Josephson phase.
The TLL parameter $g$ is equal to 1 for the noninteracting
case and smaller than 1 for repulsive interactions.
The Hamiltonian (\ref{H_0}) is the Gaussian model
describing free massless bosons with linear dispersion
$\omega=v|k|$.

Let us introduce a short-range impurity potential at $x=0$
and discuss tunneling of particles through
the potential barrier.
The scattering by the impurity potential is
described by
\begin{align}
&
\lambda_F:\!\psi^\dagger_L(0)\psi^{}_L(0) + (L\to R)\!:
+\lambda_B\!\left[\psi^\dagger_L(0)\psi^{}_R(0)
   +\mathrm{H.c.}\right]
\nonumber\\
&\qquad=
\frac{\lambda_F}{2\pi}\frac{d\phi(0)}{dx}
-\frac{\lambda_B}{\pi\alpha}\sin\phi(0).
\label{lambda_f+lambda_b}
\end{align}
The coupling constants $\lambda_F$ and $\lambda_B$ characterize
strength of forward and backward scattering by the impurity,
respectively.
The forward scattering can be absorbed by the transformation
$\psi_L\to\psi_L e^{i(\lambda_F/v_F)\Theta(x)}$
and
$\psi_R\to\psi_R e^{-i(\lambda_F/v_F)\Theta(x)}$,
and thus we will not consider it further in this section.
The backward scattering becomes a local nonlinear operator
of the $\phi$ field and is renormalized by interactions.

Since the perturbation is localized at $x=0$, we may integrate
out the fields away from $x=0$ and derive an effective
action for the field $\phi_0=\phi(x=0)$.
To this end, we write the partition function of the Gaussian
model as
\begin{equation}
Z_0[\phi_0]=\int\mathcal{D}\phi(x,\tau)\mathcal{D}\lambda(\tau)
e^{-S_0},
\label{Z_0}
\end{equation}
where $\lambda$ is an auxiliary field, and the action is
\begin{align}
S_0=&
\frac{v}{8\pi g}\int^{1/T}_0\!\!d\tau\int^\infty_{-\infty}\!\!dx
\left[\left(\frac{\partial\phi}{\partial x}\right)^2
+\frac{1}{v^2}\left(\frac{\partial\phi}{\partial\tau}\right)^2
\right]
\nonumber\\
&{}
+i\int^{1/T}_0\!\!d\tau\lambda(\tau)[\phi_0(\tau)-\phi(0,\tau)].
\end{align}
We integrate out $\phi$ first and then $\lambda$
to obtain the effective action for $\phi_0$,
\begin{equation}
S_{\mathrm{eff}}=\sum_{\omega_n}\frac{|\omega_n|}{4\pi g}
|\tilde\phi_0(i\omega_n)|^2
-\frac{\lambda_B}{\pi\alpha}\int^{1/T}_0\!\!d\tau
 \sin\phi_0(\tau),
\label{S_eff}
\end{equation}
where $\omega_n=2\pi Tn$ ($n\in\mathbb{Z}$) and
\begin{equation}
\phi_0(\tau)=
\sqrt{T}\sum_{\omega_n}e^{-i\omega_n\tau}\tilde\phi_0(i\omega_n).
\end{equation}
The first term in eq.\ (\ref{S_eff}) is reminiscent of the
Ohmic dissipation term in the Caldeira-Leggett action,\cite{leggett}
which is caused by the gapless excitations in the TLL.
If we added a ``kinetic energy'' term $m(\partial_\tau\phi_0)^2$
to regularize high-energy divergence, then our effective action
would become the same as that used in the problem of macroscopic
quantum coherence (MQC).

When $\lambda_B=0$, it is easy to calculate two-time
correlation function with the effective action, yielding
\begin{equation}
\langle e^{i\mu\phi_0(\tau)}e^{-i\mu\phi_0(0)}\rangle
\propto\tau^{-2g\mu^2}
\quad\mathrm{at}~T=0.
\label{correlator}
\end{equation}
Equation (\ref{correlator}) tells that the scaling dimension
of the backward-scattering operator $\sin\phi_0$ is $g$
at the Gaussian fixed point.
Hence, in lowest order, the scaling equation for $\lambda_B$
is given by
\begin{equation}
\frac{d\lambda_B}{dl}=(1-g)\lambda_B,
\label{dlambda_b/dl}
\end{equation}
where $dl=-d\Lambda/\Lambda$ with $\Lambda$ high-energy
cutoff.
The TLL parameter $g$, on the other hand, is not renormalized
since the local perturbation cannot affect the coupling
constant in the bulk fixed-point Hamiltonian $\mathcal{H}_0$.
The scaling equation (\ref{dlambda_b/dl}) indicates that,
as energy scale decreases,
the backscattering becomes stronger (weaker) for $g<1$ ($g>1$).
At the noninteracting point $g=1$, the coupling $\lambda_B$
is not renormalized.

When the coupling $\lambda_B$ is large, we cannot rely on
the perturbative scaling equation (\ref{dlambda_b/dl}).
In this case the field $\phi_0$ is almost always
pinned at $\phi_0=\pi/2$ (mod $2\pi$) and occasionally changes
by $\pm2\pi$.
To examine if a tunneling event, say, from $\phi_0=\pi/2$
to $\phi_0=5\pi/2$ at time $\tau_0$ is relevant or not,
we substitute $\phi_0(\tau)=\pi/2+2\pi\Theta(\tau-\tau_0)$
into the action (\ref{S_eff}),
\begin{equation}
S_\mathrm{eff}+\frac{\lambda_B}{\pi\alpha T}=
\frac{2\pi T}{g}\sum_{\omega_n}\frac{1-\cos(\omega_n\tau_0)}{|\omega_n|}
\propto\frac{2}{g}\ln\tau_0.
\label{tunnel}
\end{equation}
We can conclude from eq.\ (\ref{tunnel}) that the scaling equation for
fugacity $t$ of the tunneling is
\begin{equation}
\frac{dt}{dl}=\left(1-\frac{1}{g}\right)t.
\label{dt/dl}
\end{equation}
The tunneling is relevant (irrelevant) for $g>1$ ($g<1$).
Note the duality between eqs.\ (\ref{dlambda_b/dl}) and (\ref{dt/dl}).

\begin{figure}[tb]
\begin{center}
\begin{picture}(200,90)(0,0)
\thicklines
\put(10,10){\vector(1,0){160}}
\put(10,10){\line(0,1){70}}
\put(10,80){\line(1,0){155}}
\put(5,0){0}
\put(2,76){1}
\put(-2,40){$\mathcal{T}$}
\put(173,7){$g$}
\put(88,0){1}
\put(37,65){\vector(0,-1){2}}
\put(37,21){\vector(0,-1){2}}
\put(63,65){\vector(0,-1){2}}
\put(63,21){\vector(0,-1){2}}
\put(117,25){\vector(0,1){2}}
\put(117,67){\vector(0,1){2}}
\put(143,25){\vector(0,1){2}}
\put(143,67){\vector(0,1){2}}
\thinlines
\put(37,10){\line(0,1){70}}
\put(63,10){\line(0,1){70}}
\put(90,10){\line(0,1){70}}
\put(117,10){\line(0,1){70}}
\put(143,10){\line(0,1){70}}
\end{picture}
\end{center}
\caption{Schematic flow diagram for the transmission
probability.
There is now flow at $g=1$.}
\label{fig:flow}
\end{figure}
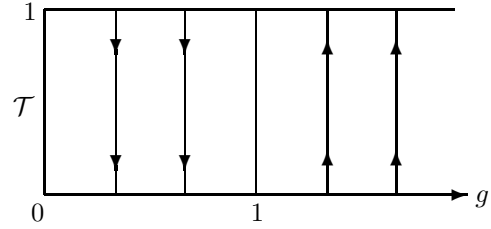

Let us denote by $\mathcal{T}$ transmission probability of a density
wave through the potential barrier.
Combining the scaling equations (\ref{dlambda_b/dl}) and (\ref{dt/dl}),
we can draw a schematic renormalization-group flow diagram\cite{kane92prl}
(Fig.\ \ref{fig:flow}).
For repulsive interactions ($g<1$) the transmission probability is
renormalized down to zero, and a density wave is perfectly reflected
in the low-energy limit.
The TLL is effectively cut into two pieces by an impenetrable barrier.
On the contrary, for attractive interactions ($g>1$) the density wave
is free to transmit through the barrier in the low-energy limit,
in agreement with quasi-long-range order of superconductivity.
At the noninteracting point $g=1$, the potential is marginal
and not renormalized at all,
in agreement with our common knowledge that
the transmission probability of a noninteracting particle can
take any value (between 0 and 1) and is a smooth function
of the potential strength.

In the MQC problem it is well known that an explicit solution 
of the problem is possible at a particular value of the
dissipation strength which corresponds to $g=1/2$ in our
problem.
At this point the scaling dimension of $\sin\phi_0$ becomes
$1/2$, indicating that this operator can be written as
a fermion operator.
To see this, let us introduce new chiral
bosonic fields\cite{affleck94}
\begin{align}
\varphi_{\pm}(x)=&
\frac{1}{\sqrt8}\left\{
\left(\frac{1}{\sqrt{g}}-\sqrt{g}\right)
[\varphi_R(x)\pm\varphi_L(-x)]
\right.\nonumber\\
&\left.\quad{}
+\left(\frac{1}{\sqrt{g}}+\sqrt{g}\right)
[\varphi_L(x)\pm\varphi_R(-x)]
\right\},
\end{align}
which satisfy
$[\varphi_\pm(x),\varphi_\pm(y)]=-i\pi\,\mathrm{sgn}(x-y)$
and $[\varphi_+(x),\varphi_-(y)]=i\pi$.
The total Hamiltonian (\ref{H_0})+(\ref{lambda_f+lambda_b})
can then be written as $\mathcal{H}_F+\mathcal{H}_B$, where 
\begin{align}
\mathcal{H}_F&=
\frac{v}{4\pi}\int^\infty_{-\infty}\!\!
\left(\frac{d\varphi_-}{dx}\right)^2\!dx
+\frac{\lambda_F}{\pi}\sqrt{\frac{g}{2}}\frac{d\varphi_-(0)}{dx},
\label{H_F}
\\
\mathcal{H}_B&=
\frac{v}{4\pi}\int^\infty_{-\infty}\!\!
\left(\frac{d\varphi_+}{dx}\right)^2\!dx
-\frac{\lambda_B}{\pi\alpha}\sin[\sqrt{2g}\varphi_+(0)],
\label{H_B}
\end{align}
and $[\mathcal{H}_F,\mathcal{H}_B]=0$.
In this way the forward and backward scattering processes can
be separated into two independent problems.
At $g=1/2$ we fermionize the vertex operator
\begin{equation}
\frac{e^{-i\varphi_+(x)}}{\sqrt{2\pi\alpha}}
=\eta\psi_+(x)
\end{equation}
with a chiral fermion field $\psi_+(x)$ and a Majorana fermion
$\eta$, which satisfy $\{\psi_+(x),\eta\}=0$ and $\eta^2=1$.
The Hamiltonian $\mathcal{H}_B$ can then be
refermionized as\cite{matveev95}
\begin{align}
\mathcal{H}_B=& \,
iv\int^\infty_{-\infty}\psi_+(x)\frac{d}{dx}\psi_+(x)dx
\nonumber\\
&{}
+i\frac{\lambda_B}{\sqrt{2\pi\alpha}}
\left[\eta\psi^{}_+(0)+\psi^\dagger_+(0)\eta\right],
\label{H_B2}
\end{align}
which is a quadratic Hamiltonian and easy to diagonalize.
The transmission probability $\mathcal{T}$ is
obtained\cite{kane92prb,matveev95} as
\begin{equation}
\mathcal{T}_{g=1/2}=
\int^\infty_{-\infty}dE\frac{E^2}{E^2+\Gamma^2}
\left(-\frac{d}{dE}\right)\frac{1}{e^{E/T}+1},
\end{equation}
where $\Gamma=\lambda^2/\pi\alpha v_F$.

In the low-energy limit the Hamiltonian (\ref{H_B2}) is
equivalent\cite{matveev95}
to the one obtained by Emery and Kivelson\cite{emerykivelson}
for the Toulouse limit of the two-channel Kondo model.
A ``half'' of the impurity spin ($S=1/2$) becomes the Majorana
fermion $\eta$ and the other ``half'' is decoupled from the
Fermi bath, yielding the anomalous residual entropy in the
two-channel Kondo problem.

Even away from the special point $g=\frac12$, it is possible
to solve $\mathcal{H}_B$ exactly with the help of the
Bethe ansatz solution to the boundary sine-Gordon model.\cite{fendley}
However, the calculation and resulting formulas are
not as simple as in the $g=\frac12$ case.

\subsection{spinful case}

A TLL of electrons in a quantum wire has internal degree(s) of
freedom: spin (and ``flavor'' or band index in nanotubes),
and thus it is important to study the quantum impurity problem
with spins.
Indeed, the spinless model was generalized to spinful electrons
immediately\cite{furusaki93} after the spinless
case was studied.\cite{kane92prl}
Let us briefly review the single-impurity problem for
a spinful TLL.

The model we consider is a simple generalization of the spinless
case.
The field operator of right- and left-going electrons with
spin $\sigma$ are written as
\begin{subequations}
\begin{align}
\psi_{R\sigma}(x)&=
\frac{\eta_\sigma}{\sqrt{2\pi\alpha}}e^{i\varphi_{R\sigma}(x)},
\\
\psi_{L\sigma}(x)&=
\frac{\eta_\sigma}{\sqrt{2\pi\alpha}}e^{-i\varphi_{L\sigma}(x)},
\end{align}
\end{subequations}
where the bosonic fields $\varphi_{R\sigma}$ and
$\varphi_{L\sigma}$ satisfy
\begin{subequations}
\begin{align}
[\varphi_{R\sigma}(x),\varphi_{L\sigma'}(y)]&=
i\pi\delta_{\sigma,\sigma'}\mathrm{sgn}(x-y),\\
[\varphi_{L\sigma}(x),\varphi_{L\sigma'}(y)]&=
-i\pi\delta_{\sigma,\sigma'}\mathrm{sgn}(x-y),\\
[\varphi_{R\sigma}(x),\varphi_{L\sigma'}(y)]&=
i\pi\delta_{\sigma,\sigma'},
\end{align}
\end{subequations}
and the Klein factors (Majorana fermions) $\eta_\sigma$
obeying
\begin{equation}
\{\eta_\sigma,\eta_{\sigma'}\}=2\delta_{\sigma,\sigma'}
\end{equation}
have been introduced to respect the anticommutation relation
between electron fields with antiparallel spins.

Since spin and charge degrees of freedom are separated in
a spinful TLL, the low-energy effective theory is given by
a sum of two independent Gaussian models.
The Hamiltonian of the spinful TLL is given by
\begin{align}
\mathcal{H}_s=&
\frac{v_\rho}{4\pi}\int^\infty_{-\infty}\! dx
\left[
\frac{1}{K_\rho}\left(\frac{d\phi_\rho}{dx}\right)^2
+K_\rho\left(\frac{d\theta_\rho}{dx}\right)^2
\right]
\nonumber\\
&
+
\frac{v_\sigma}{4\pi}\int^\infty_{-\infty}\! dx
\left[
\frac{1}{K_\sigma}\left(\frac{d\phi_\sigma}{dx}\right)^2
+K_\sigma\left(\frac{d\theta_\sigma}{dx}\right)^2
\right],
\label{H_s}
\end{align}
where
\begin{subequations}
\begin{align}
\phi_\rho(x)&\!=\frac{1}{2}
[\varphi_{L\uparrow}(x)+\varphi_{L\downarrow}(x)
+\varphi_{R\uparrow}(x)+\varphi_{R\downarrow}(x)],
\\
\theta_\rho(x)&\!=\frac{1}{2}
[\varphi_{L\uparrow}(x)+\varphi_{L\downarrow}(x)
-\varphi_{R\uparrow}(x)-\varphi_{R\downarrow}(x)],
\\
\phi_\sigma(x)&\!=\frac{1}{2}
[\varphi_{L\uparrow}(x)-\varphi_{L\downarrow}(x)
+\varphi_{R\uparrow}(x)-\varphi_{R\downarrow}(x)],
\\
\theta_\sigma(x)&\!=\frac{1}{2}
[\varphi_{L\uparrow}(x)-\varphi_{L\downarrow}(x)
-\varphi_{R\uparrow}(x)+\varphi_{R\downarrow}(x)].
\end{align}
\end{subequations}
The parameters $K_\rho$ and $K_\sigma$ are the TLL parameters
controlling charge and spin sectors, respectively.
Roughly speaking, they are smaller (larger) than unity when
density-density interactions are repulsive (attractive).
For noninteracting electrons $K_\rho=K_\sigma=1$.
For example, in the Hubbard model with on-site repulsion $U$
the parameter $K_\rho$ decreases from 1 to $\frac12$ as $U$
increases from 0 to infinity.
With longer-range interactions $K_\rho$ can take a value
smaller than $\frac12$.
It is also important to note that, when the Hamiltonian
has SU(2) spin rotation symmetry, $K_\sigma$ is renormalized
to 1 in the low-energy limit.
The approach to the fixed-point value $K_\sigma^*=1$ is
logarithmic and is controlled by the bulk leading irrelevant
operator $\cos(2\phi_\sigma)$, omitted in eq.\ (\ref{H_s}),
arising from backward scattering
of electrons with antiparallel spins.
Although this effect is important for quantitative
analysis at finite temperatures, we shall adopt the
fixed-point Hamiltonian (\ref{H_s}) for our discussion
of the impurity model in the low-energy limit.
We will come back to this issue at the end of this paper.

The backward scattering off the static impurity at $x=0$
is caused by the impurity Hamiltonian
\begin{align}
\mathcal{H}'&=
\lambda_B\sum_\sigma\left[
\psi^\dagger_{L\sigma}(0)\psi^{}_{R\sigma}(0)
+\psi^\dagger_{R\sigma}(0)\psi^{}_{L\sigma}(0)
\right]
\nonumber\\
&=
-\frac{2\lambda_B}{\pi\alpha}\sin[\phi_\rho(0)]\cos[\phi_\sigma(0)].
\label{H'}
\end{align}
Note that there is no spinflip scattering.
In the bosonic picture the effect of the impurity backscattering
is to pin both charge and spin density fields $\phi_\rho$ and
$\phi_\sigma$.
The competition between the kinetic energy (\ref{H_s})
of density waves and the pinning potential (\ref{H'}) can be
examined through perturbative renormalization group in the same
way as in the spinless model.
On the one hand, the scaling equation for backscattering $\lambda_B$,
corresponding to eq.\ (\ref{dlambda_b/dl}),
reads
\begin{equation}
\frac{d\lambda_B}{dl}=\frac{1}{2}(2-K_\rho-K_\sigma)\lambda_B.
\label{dlambda_B/dl2}
\end{equation}
On the other hand, the scaling equation for fugacity $t$ of
single-electron tunneling is generalized from eq.\ (\ref{dt/dl}) to
\begin{equation}
\frac{dt}{dl}=
\frac{1}{2}\left(2-\frac{1}{K_\rho}-\frac{1}{K_\sigma}\right)t.
\label{dt/dl2}
\end{equation}

In the SU(2) symmetric case where $K_\sigma=1$, the scaling
equations (\ref{dlambda_B/dl2}) and (\ref{dt/dl2}) are
essentially the same as in the spinless case.
Hence the flow diagram (Fig.\ \ref{fig:flow}) applies.
For repulsive interactions the TLL wire is cut into two
semi-infinite wires, while the weak link is healed for
attractive interactions.

If we allow $K_\rho$ and $K_\sigma$ to be free
parameters,\cite{furusaki93} the flow diagram becomes much
richer.\cite{furusaki93,kane92prb,wong,oshikawa}
For example, when $K_\sigma>1$ and
$2-K_\sigma<K_\rho<K_\sigma/(2K_\sigma-1)$,
both $\lambda_B$ and $t$ are renormalized to smaller values,
implying that there should be a line (surface) of critical points
in the intermediate-coupling regime, where the transmission
probability $\mathcal{T}$ takes a nontrivial value.
It remains to be understood what kind of boundary
conditions should be obeyed by the bosonic fields $\phi_\rho$
and $\phi_\sigma$ at the critical point.

\begin{figure}[tb]
\begin{center}
\begin{picture}(200,90)(0,0)
\thicklines
\put(10,10){\vector(1,0){160}}
\put(10,10){\line(0,1){70}}
\put(10,80){\line(1,0){155}}
\put(5,0){0}
\put(2,76){1}
\put(-2,40){$\mathcal{T}$}
\put(173,7){$K_\rho$}
\put(33,68){\vector(0,-1){2}}
\put(33,18){\vector(0,-1){2}}
\put(65,72){\vector(0,1){2}}
\put(60,18){\vector(0,-1){2}}
\put(85,14){\vector(0,-1){2}}
\put(88,72){\vector(0,1){2}}
\put(122,22){\vector(0,1){2}}
\put(115,72){\vector(0,1){2}}
\put(143,22){\vector(0,1){2}}
\put(143,72){\vector(0,1){2}}
\put(45,80){\line(1,-1){70}}
\put(115,10){\line(-1,1){20}}
\thinlines
\put(33,10){\line(0,1){20}}
\put(33,80){\line(0,-1){20}}
\put(60,10){\line(0,1){20}}
\put(65,80){\line(0,-1){15}}
\put(85,10){\line(0,1){15}}
\put(88,80){\line(0,-1){18}}
\put(115,80){\line(0,-1){20}}
\put(122,10){\line(0,1){20}}
\put(143,10){\line(0,1){20}}
\put(143,80){\line(0,-1){20}}
\end{picture}
\end{center}
\caption{Schematic flow diagram for the transmission
probability at $K_\sigma>1$.
The line of critical points is drawn as a straight
line for simplicity.}
\label{fig:flow2}
\end{figure}
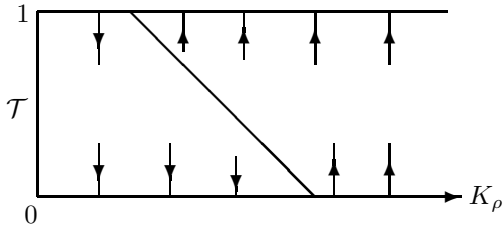

We emphasize that the simplest duality picture
(Fig.~\ref{fig:flow})
holds only for the spinless case where there is essentially
only one bosonic field $\phi$.
Once we generalize the model to include more than one bosonic
fields as in the spinful case, the problem becomes
rather nontrivial.
It is a challenging problem to find conformally invariant
boundary conditions, other than the
Dirichlet and Neumann boundary conditions we saw in the spinless case.
Recent theoretical attempts in this direction can be found
in ref.~\citen{oshikawa,yi,chamon}.
A similar problem also arises in the study of quantum and dissipative
Josephson junctions.\cite{refael}

\section{Orthogonality catastrophe}

The orthogonality catastrophe\cite{anderson}
and Fermi-edge singularities\cite{nozieres} have been
milestones in the theory of quantum impurities in Fermi liquids.
Thus it would be meaningful to briefly review the analogous
problem for a TLL.

For simplicity we discuss the orthogonality catastrophe
in the spinless case.
Let $|0\rangle$ denote the ground state of a clean TLL
and $|\lambda\rangle$ the ground state of a TLL with
the scattering potential (\ref{lambda_f+lambda_b}) at $x=0$.
The quantity of our interest is the overlap integral
$|\langle0|\lambda\rangle|^2$.
The overlap is expected to vanish
in the thermodynamic limit
as $L^{-\gamma}$, where $L$ is the length of the one-dimensional
system.
Since the TLL Hamiltonian (\ref{H_0}) plus the potential
(\ref{lambda_f+lambda_b}) can be separated
into two commuting parts $\mathcal{H}_F$ and $\mathcal{H}_B$,
the orthogonality problem can be discussed separately
for $\mathcal{H}_F$ and $\mathcal{H}_B$.
Accordingly the exponent $\gamma$ is written as
$\gamma_F+\gamma_B$.

The forward-scattering exponent $\gamma_F$ can be found
easily by a unitary transformation.\cite{ogawa,dkklee}
Using the commutation relation
\begin{equation}
\frac{\partial}{\partial y}
[\varphi_-(x),\varphi_-(y)]=2\pi i\delta(x-y)
\end{equation}
we transform the Hamiltonian $\mathcal{H}_F$ as
\begin{equation}
U\mathcal{H}_F U^\dagger
=\frac{v}{4\pi}\int dx
\left(\frac{d\varphi_-}{dx}\right)^2
+\mathrm{const},
\label{UHU}
\end{equation}
where
\begin{equation}
U=\exp\left[
i\frac{\lambda_F}{\pi v}\sqrt{\frac{g}{2}}\varphi_-(0)
\right].
\end{equation}
The overlap can then be calculated as
$|\langle0|U|0 \rangle|^2\propto L^{-\gamma_F}$ with
\begin{equation}
\gamma_F=2g\left(\frac{\lambda_F}{2\pi v}\right)^2.
\end{equation}

The overlap integral in the even ($\varphi_+$) sector
can be found from the following consideration to take
a universal value independent of both mutual interaction
strength and impurity potential. 
In \S2.1 we have seen that the renormalization-group flow
of the backscattering potential is different for attractive
and repulsive interactions.
At $g>1$ the backscattering potential is renormalized to
zero in the low-energy limit.
This means that there should remain a finite overlap
integral even in the thermodynamic limit; $\gamma_B=0$.
At $g<1$ the backscattering potential grows and eventually
the TLL is effectively cut into two decoupled pieces.
This means that, roughly speaking, a density wave of even parity
changes from $\cos(kx)$ at $\lambda_B=0$ to $\sin|kx|$ for
any $\lambda_B\ne0$ as $k\to0$.
The phase shift at $k\to0$ is thus $\pi/2$ for any $g<1$ and
$\lambda_B\ne0$, and we can expect that the exponent $\gamma_B$
should take a universal value.
The exponent can be obtained in various
ways.\cite{prokofev,gogolin,kane94,affleck94,komnik97,furusaki97}
Since the relevant sine potential strongly pins the $\varphi_+$
field, one may replace the potential with a local mass term,
$m\varphi_+^2(0)$.
Now that the Hamiltonian is quadratic, it is not difficult to
compute the overlap.\cite{prokofev,gogolin}
Other approaches include renormalization-group analysis in
the weak-interaction limit,\cite{kane94} boundary conformal
field theory\cite{affleck94} and
refermionization\cite{komnik97,furusaki97} at $g=\frac12$.
All these methods give
\begin{equation}
\gamma_B=\frac{1}{8}.
\label{1/8}
\end{equation}
One way to understand this result is use free-fermion picture
in the weak-interaction limit.\cite{kane94}
It is known\cite{anderson} that the orthogonality exponent $\gamma$
is a sum of odd and even wave functions,
$\gamma=(\delta_o/\pi)^2+(\delta_e/\pi)^2$, where
$\delta_o$ and $\delta_e$ are phase shifts in the wave functions,
\begin{subequations}
\begin{align}
\psi_o(x)&=\sin(kx+\delta_o\mathrm{sgn}x),\\
\psi_e(x)&=\cos(kx+\delta_e\mathrm{sgn}x).
\end{align}
\end{subequations}
The scattering state,
\begin{equation}
\tilde\psi(x)=
\begin{cases}
e^{ikx}+\tilde re^{-ikx},& x<0,\cr
\tilde te^{ikx},&x>0,\cr
\end{cases}
\end{equation}
is related to $\psi_o$ and $\psi_e$ by
\begin{equation}
\tilde\psi(x)=ie^{i\delta_o}\psi_o(x)+e^{i\delta_e}\psi_e(x),
\end{equation}
from which we find the transmission amplitude
$\tilde t=e^{i\delta_+}\cos\delta_-$
with $\delta_\pm=\delta_e\pm\delta_o$.
As is obvious from the discussion below eq.\ (\ref{lambda_f+lambda_b}),
the phase shift $\delta_+$ is proportional to $\lambda_F$.
Since $\mathcal{T}=|\tilde t|^2=\cos^2\delta_-$, the phase shift
$\delta_-$ is controlled by $\lambda_B$ and approaches
$\pi/2$ in the low-energy limit (Fig.~\ref{fig:flow}).
The exponent $\gamma_B$ is then given by
$(\delta_-/\pi)^2/2=\frac18$.

Another interpretation to eq.\ (\ref{1/8}) is that, in the
refermionized Hamiltonian $\mathcal{H}_B$ at $g=\frac12$,
the impurity (or $\eta$) is coupled only to
$\psi_+^{}(0)-\psi_+^\dagger(0)$, and the
other half degree of freedom $\psi_+^{}+\psi_+^\dagger$ is
decoupled.
Hence only half of the fermions $\psi_+$ have the phase shift
$\delta=\pi/2$, leading to the exponent $(\delta/\pi)^2/2=\frac18$.

\section{Magnetic impurity}

Finally, we turn to the Kondo effect of one-dimensional
repulsively interacting electrons.
Suppose that there is an impurity spin ($S=\frac12$)
at the origin.
As we have seen in the previous sections, in one dimension we
need to distinguish forward and backward scattering by
the impurity, unless the impurity spin is at the boundary.
We thus consider two kinds of Kondo exchange couplings:
forward Kondo scattering $J_F$ and backward Kondo scattering $J_B$.
The Kondo scattering is described by
\begin{align}
\mathcal{H}_J=&
\frac{J_F}{2}\mib{S}\cdot\!\left[
\psi^\dagger_{R\alpha}(0)\mib{\sigma}_{\alpha\beta}
\psi^{}_{R\beta}(0)
+\psi^\dagger_{L\alpha}(0)\mib{\sigma}_{\alpha\beta}
\psi^{}_{L\beta}(0)
\right]
\nonumber\\
&
+\frac{J_B}{2}\mib{S}\cdot\!\left[
\psi^\dagger_{R\alpha}(0)\mib{\sigma}_{\alpha\beta}^{}
\psi^{}_{L\beta}(0)
+\psi^\dagger_{L\alpha}(0)\mib{\sigma}_{\alpha\beta}^{}
\psi^{}_{R\beta}(0)
\right],
\label{H_J}
\end{align}
where $\mib{\sigma}=(\sigma_x,\sigma_y,\sigma_z)$ are
Pauli matrices.

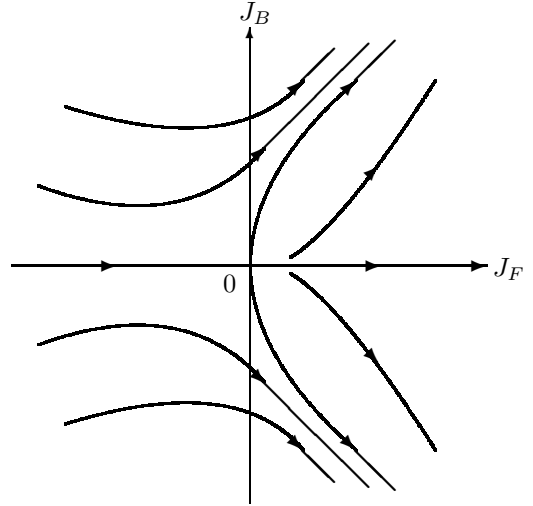
\begin{figure}[tb]
\begin{center}
\begin{picture}(200,200)(0,0)
\put(90,90){0}
\put(100,10){\vector(0,1){180}}
\put(96,193){$J_B$}
\put(192,96){$J_F$}
\thicklines
\put(10,100){\vector(1,0){180}}
\put(30,100){\vector(1,0){20}}
\put(130,100){\vector(1,0){20}}
\qbezier(140,30)(60,100)(140,170)
\put(140,170){\line(1,1){15}}
\put(138,168){\vector(1,1){2}}
\put(140,30){\line(1,-1){15}}
\put(138,32){\vector(1,-1){2}}
\qbezier(30,160)(95,140)(120,170)
\put(120,170){\line(1,1){12}}
\put(118,168){\vector(1,1){2}}
\qbezier(30,40)(95,60)(120,30)
\put(120,30){\line(1,-1){12}}
\put(118,32){\vector(1,-1){2}}
\qbezier(20,130)(75,110)(105,144)
\put(105,144){\line(1,1){40}}
\put(104,143){\vector(1,1){2}}
\qbezier(20,70)(75,90)(105,56)
\put(105,56){\line(1,-1){40}}
\put(104,57){\vector(1,-1){2}}
\qbezier(115,103)(135,115)(170,170)
\put(147,136){\vector(1,1){2}}
\qbezier(115,97)(135,85)(170,30)
\put(147,65){\vector(1,-1){2}}
\end{picture}
\end{center}
\caption{Schematic flow diagram for the Kondo couplings
for $K_\rho<1$ and $K_\sigma=1$.}
\label{fig:flow3}
\end{figure}

We assume that the bulk TLL is described by the
Hamiltonian $\mathcal{H}_s$ (\ref{H_s}), where
the SU(2) spin rotation symmetry implies $K_\sigma=1$
and $K_\rho$ is less than 1 (repulsive interaction).
The Kondo scattering (\ref{H_J}) can be bosonized as in \S2.2.
The scaling dimension of magnetic impurity scattering is the
same as that of nonmagnetic impurity scattering.
Thus, the scaling dimension of $J_F$ is one, whereas
$J_B$ has dimension $(1+K_\rho)/2$.
Up to one-loop order, renormalization-goup equations can be
easily derived using, for example, the poor-man's scaling
method.\cite{poor-man}
The scaling equations read
\begin{subequations}
\begin{align}
\frac{dJ_F}{dl}&=\frac{1}{2\pi v}(J_F^2+J_B^2),
\label{dJ_F/dl}\\
\frac{dJ_B}{dl}&=\frac{1}{2}(1-K_\rho)J_B+\frac{1}{\pi v}J_FJ_B.
\label{dJ_B/dl}
\end{align}
\end{subequations}
A schematic flow diagram is shown in Fig.\ \ref{fig:flow3}.

The flow diagram tells that the trivial fixed point $J_F=J_B=0$
becomes unstable with infinitesimal $J_B$, because the backward
scattering is relevant for $K_\rho<1$.
This leads to a somewhat surprising conclusion that the Kondo couplings
are renormalized towards strong couplings, no matter whether
the Kondo couplings are antiferromagnetic or ferromagnetic.
This should be contrasted with the standard Fermi liquid case
in which the Kondo coupling is renormalized to the strong-coupling
regime only if it is antiferromagnetic.
Another point to note is that the Kondo temperature depends
on the Kondo coupling constant not exponentially but algebraically
in TLLs.

The one-loop scaling equations (\ref{dJ_F/dl}) and (\ref{dJ_B/dl})
suggest three fixed points besides the trivial one, $(J_F,J_B)=(0,0)$,
mentioned above.
These three are $(J_F,J_B)=(+\infty,+\infty)$, $(+\infty,0)$
and $(+\infty,-\infty)$.

\begin{figure}
\begin{center}
\begin{picture}(200,100)(0,0)
\put(90,70){\line(1,2){10}}
\put(110,70){\line(-1,2){10}}
\put(90,77){\vector(0,-1){14}}
\put(110,77){\vector(0,-1){14}}
\put(89,80){$J$}
\put(106,80){$J$}
\thicklines
\put(0,70){\line(1,0){200}}
\multiput(10,70)(20,0){10}{\circle*{5}}
\put(100,80){\vector(0,1){16}}
\put(98,40){$\Downarrow$}
\put(0,15){\line(1,0){70}}
\multiput(10,15)(20,0){4}{\circle*{5}}
\put(130,15){\line(1,0){70}}
\multiput(130,15)(20,0){4}{\circle*{5}}
\put(100,23){\vector(0,-1){16}}
\thinlines
\multiput(74,15)(12,0){5}{\line(1,0){6}}
\put(82,5){$J'$}
\put(112,5){$J'$}
\end{picture}
\end{center}
\caption{Lattice model in which an impurity spin is coupled
antiferromagnetically to two lattice sites.}
\label{fig:J_B=0}
\end{figure}
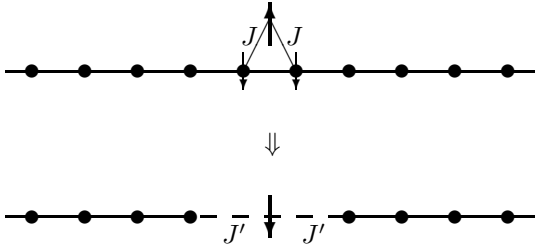

Let us first discuss the case $J_B=0$.
In this case the charge and spin sectors are decoupled in
the bosonized Hamiltonian $H_s+H_J$.
The impurity spin is interacting with $\phi_s$ and $\theta_s$
only, and the Hamiltonian of the spin sector is equivalent
to that of the two-channel
Kondo problem.\cite{affleckludwig}
Here the right- and left-going electrons correspond to two
channels.
As is well known, the ferromagnetic coupling $J_F>0$ is renormalized
to zero, while the antiferromagnetic coupling flows to
the strong-coupling fixed point of the two-channel Kondo model.
In the latter case the specific heat and the spin susceptibility
acquire anomalous logarithmic contributions due to a local
leading irrelevant operator with dimension $\frac32$ at the
strong-coupling fixed point:
\begin{equation}
\delta C\propto T\ln(T_K/T),\qquad
\delta\chi\propto\ln(T_K/T).
\end{equation}
Since the charge excitations have the velocity $v_\rho$ which
is different from the spin velocity $v_\sigma$, the Wilson ratio
is slightly modified\cite{frojdh} from the standard result $\frac83$ to
\begin{equation}
R_W=\frac{4}{3}\left(1+\frac{v_\sigma}{v_\rho}\right).
\end{equation}
A simple lattice model realizing this two-channel Kondo physics
is depicted in Fig.\ \ref{fig:J_B=0}, where an impurity spin
$S=\frac12$ is coupled to two neighboring sites
with equal antiferromagnetic Kondo coupling $J$.
When the electron density is at half filling, one can easily show
that the backward scattering vanishes in the continuum
limit.\cite{furusaki94,frojdh}
Since the charge sector is gapped at half filling, the model
is essentially the same as the antiferromagnetic Heisenberg
chain with the impurity spin.\cite{eggert,clarke}

\begin{figure}
\begin{center}
\begin{picture}(200,100)(0,0)
\put(20,90){(a)}
\put(80,70){\line(1,0){40}}
\put(100,77){\vector(0,-1){14}}
\thicklines
\put(10,70){\line(1,0){70}}
\put(190,70){\line(-1,0){70}}
\multiput(20,70)(20,0){4}{\circle*{5}}
\multiput(120,70)(20,0){4}{\circle*{5}}
\put(100,80){\vector(0,1){16}}
\put(100,79){\oval(18,46)}
\put(113,88){$S=0$}
\put(78,59){$l$}
\put(117,60){$r$}
\thinlines
\put(20,33){(b)}
\put(60,15){\line(1,0){80}}
\put(80,22){\vector(0,-1){14}}
\put(120,22){\vector(0,-1){14}}
\put(100,8){\vector(0,1){14}}
\thicklines
\put(10,15){\line(1,0){50}}
\multiput(20,15)(20,0){3}{\circle*{5}}
\put(140,15){\line(1,0){50}}
\multiput(140,15)(20,0){3}{\circle*{5}}
\put(100,25){\vector(0,1){16}}
\put(100,22){\oval(54,50)}
\put(129,33){$S=0$}
\put(58,4){$l$}
\put(137,5){$r$}
\end{picture}
\end{center}
\caption{Schematic pictures of the strong-coupling fixed
points for (a) antiferromagnetic and (b) ferromagnetic
Kondo couplings.
}
\label{fig:strong}
\end{figure}
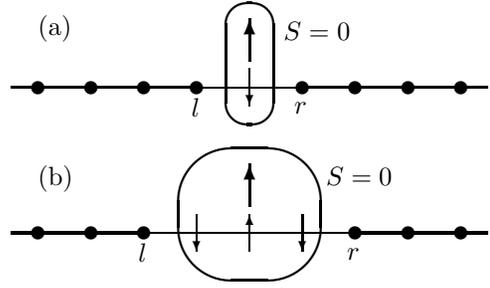

We now turn our attention to the generic case $J_FJ_B\ne0$.
Since the Kondo couplings always flow towards strong couplings
(Fig.\ \ref{fig:flow3}), it is natural to assume that in the
ground state the impurity spin is completely screened as in
the standard Kondo effect in Fermi liquids (an exception is
the two-channel Kondo case discussed above, which requires
the condition $J_B=0$).
We can thus draw schematic strong-coupling pictures of the
stable fixed points for (a) antiferromagnetic and (b) ferromagnetic
Kondo couplings; see Fig.\ \ref{fig:strong}.
The ground state consists essentially of two semi-infinite TLLs
and a singlet in between them.
Since the isolated singlet is expected to have a finite energy gap to
excited states, these pictures suggest that the low-energy
effective theory is given by two decoupled semi-infinite TLLs
plus residual perturbations.
Before concluding that these pictures are indeed correct, we have to
examine whether the residual interactions are irrelevant at the
fixed points.
Local operators that could be generated during the renormalization
group transformation
are the local potential operator at the two ends of the TLLs
and the single-electron tunneling between the TLLs.
They can be generated by virtual breaking of the singlet due to
electron hopping into or out of the singlet.
Denoting the end sites of the left and right semi-infinite TLLs
by $l$ and $r$, respectively, we can write these operators as
$\psi^\dagger_\sigma(l)\psi^{}_\sigma(l)
+\psi^\dagger_\sigma(r)\psi^{}_\sigma(r)$
and
$\psi^\dagger_\sigma(l)\psi^{}_\sigma(r)
+\psi^\dagger_\sigma(r)\psi^{}_\sigma(l)$,
where summation over the spin index $\sigma$ is assumed.
The former is exactly marginal and can lead to a shift of the
ground state energy.
The latter operator is equivalent to the single-electron tunneling
discussed in \S2.2.
It has scaling dimension $(K_\rho^{-1}+1)/2$ at $K_\sigma=1$
and is irrelevant in the repulsively interacting electrons ($K_\rho<1$).
We can thus safely conclude that the strong-coupling fixed points
are basically two decoupled semi-infinite TLLs, and the impurity
spin is completely screened and disappears from the low-energy theory.
Treating the tunneling operator as a perturbation, we can compute
the impurity contributions to the specific heat and the spin
susceptibility at low temperatures.
The results are\cite{furusaki94}
\begin{align}
\delta C&=c_1(K_\rho-1)^2T^{1/K_\rho-1}+O(T),
\\
\delta\chi&=c_2+O(T^2),
\end{align}
where $c_1$ and $c_2$ are positive constants.
When $\frac12<K_\rho<1$, which is the case in the Hubbard model,
the low-temperature specific heat has the anomalous power-law
contribution as a leading term.
This result is confirmed by the boundary conformal field theory
analysis\cite{frojdh,durganandini} as well as by a quantum Monte Carlo
calculation.\cite{egger}
The XXZ spin chain with an extra impurity spin also shows
a similar temperature dependence.\cite{xxz}
If the single-electron tunneling operator is not allowed by
symmetry (particle-hole symmetry), then $c_1=0$ and the
low-temperature behavior is Fermi-liquid like.
This is probably what is happening in the solvable toy model studied
by Schiller and Ingersent.\cite{schiller}

One can also think of the possibility of having a static
impurity potential in addition to the magnetic impurity.
This is indeed what one would find if one takes the
asymmetric Anderson model.
Fabrizio and Gogolin\cite{fabrizio} argued that, if $K_\rho<\frac12$
and if the static potential is sufficiently strong,
then a situation similar to the one drawn in Fig.~\ref{fig:J_B=0}
occurs and the two-channel Kondo physics is realized.
Quantum Monte Carlo calculations\cite{egger} have obtained results
that are consistent with this scenario.

Recent studies have shown that this is not the whole
story.\cite{fujimoto,furusaki04}
As we have noted earlier, the low-energy theory of a spinful
TLL is the Gaussian model $H_s$ perturbed by a marginally
irrelevant operator $\cos(2\phi_\sigma)$.
Even though it is renormalized to zero in the low-energy limit,
it should be included in the calculations of finite-temperature
quantities like the specific heat and the susceptibility.
At the strong-coupling fixed point (Fig.~\ref{fig:strong}) where
the impurity spin is completely screened, the field $\phi_\sigma$
obeys a Dirichlet boundary condition at the end sites.
As a consequence the first-order perturbation
$\langle\cos(2\phi_\sigma)\rangle$
gives a nonvanishing contribution to the free energy.
That contribution comes from the region localized near the
end sites within the distance of order $v_\sigma/T$.
This boundary contribution turned out to give
leading contribution in
$\delta C$ and $\delta\chi$,
\begin{equation}
\delta C=\frac{1}{2[\ln(T_0/T)]^2},
\qquad
\delta\chi=\frac{1}{12T\ln(T_0/T)},
\label{boundary}
\end{equation}
where $T_0$ is the ``Kondo'' temperature for the bulk marginally
irrelevant operator, while the coefficients $\frac12$ and $\frac{1}{12}$
are universal.
The low-temperature behavior (\ref{boundary}) should be easily observed
when the impurity potential is sufficiently strong so that the
strong-coupling fixed point is already reached at $T_0$.

\section{Concluding remarks}

As we have seen in this paper, the quantum impurity problems in TLLs
share many interesting features with the Kondo physics.
Some of the theoretical predictions for the simplest case (a static
impurity) have been confirmed by tunneling experiments on quantum Hall
edge states and carbon nanotubes.
Experimental realizations of dynamical impurities are still to be seen
in the future.
Rapid advances in nanoscience might soon give us such cases.

\section*{Acknowledgment}

I am grateful to N.\ Nagaosa, K.\ A.\ Matveev
and T.\ Hikihara for fruitful collaborations
which helped me understand the subjects discussed in this paper.
I also appreciate the hospitality of the Aspen Center for Physics
and of Korea Institute for Advanced Study
where part of this paper was completed.
This work was partially supported by a Grant-in-Aid for
Scientific Research from the Ministry of Education, Culture,
Sports, Science and Technology (No.\ 16GS0219).

\end{document}